\documentclass[eqsecnum,preprint,prd,aps,nofootinbib]{revtex4}
\usepackage{amsmath}
\newcommand{\be}{\begin{equation}}
\newcommand{\ee}{\end{equation}}
\newcommand{\ba}{\begin{eqnarray}}
\newcommand{\ea}{\end{eqnarray}}
\newcommand{\bc}{\begin{center}}
\newcommand{\ec}{\end{center}}

\def\lira{\lambda_\ast}
\def\gira{g_\ast}
\def\ngf{g_\ast\lambda_\ast}
\def\mt{{\cal M}}
\def\gu{{h_1}}
\def\lu{h_{2}}
\def\cchi{\sqrt{8\pi}g_\ast}
\begin{document}
\begin{center}
\bibliographystyle{article}

{\Large \textsc{The Accelerated expansion of the Universe as a 
crossover phenomenon}}

\end{center}
\vspace{0.4cm}

%\vspace*{0.6cm}

\date{\today}

\author{Alfio Bonanno,$^{1,2}$ \thanks{%
Electronic address: abo@ct.astro.it} Giampiero Esposito$^{3,4}$ \thanks{%
Electronic address: giampiero.esposito@na.infn.it} Claudio Rubano$^{4,3}$
\thanks{%
Electronic address: claudio.rubano@na.infn.it} and Paolo Scudellaro$^{4,3}$
\thanks{%
Electronic address: scud@na.infn.it}}
\affiliation{${\ }^{1}$Osservatorio Astrofisico, Via S. Sofia 78, 95123 Catania, Italy\\
${\ }^{2}$Istituto Nazionale di Fisica Nucleare, Sezione di Catania,\\
Corso Italia 57, 95129 Catania, Italy\\
${\ }^{3}$Istituto Nazionale di Fisica Nucleare, Sezione di Napoli,\\
Complesso Universitario di Monte S. Angelo, Via Cintia, Edificio N', 80126
Napoli, Italy\\
${\ }^{4}$Dipartimento di Scienze Fisiche, Complesso Universitario di Monte
S. Angelo,\\
Via Cintia, Edificio N', 80126 Napoli, Italy}

\begin{abstract}
We show that the accelerated expansion of the Universe can be 
viewed as a crossover phenomenon where
the Newton constant and the Cosmological constant are actually scaling 
operators, dynamically evolving 
in the attraction basin of a non-Gaussian infrared fixed point, 
whose existence has been recently discussed.   
By linearization of the renormalized flow it is possible to 
evaluate the critical 
exponents, and it turns out that the approach to the fixed point is 
ruled by a marginal and a relevant direction. A smooth transition between the
standard Friedmann--Lemaitre--Robertson--Walker (FLRW) 
cosmology and the observed accelerated expansion is then obtained,
so that $\Omega_M \approx \Omega_\Lambda$  at late times.
\end{abstract}
\maketitle
\bigskip
\vspace{2cm}

\section{Introduction}

One of the most important questions in modern 
Cosmology is to understand the origin 
of the accelerated expansion of the Universe 
\cite{Amen03}. An original approach 
to this problem suggests the possibility that the cosmological dynamics is 
generated by strong ``renormalization group induced'' 
quantum effects which would drive
the (dimensionless) cosmological ``constant'' $\lambda(k)$ 
and Newton ``constant'' $g(k)$ 
to an infrared attractive non-Gaussian fixed point 
\cite{Bona02a,Reut02a,Reut04a}.

This hypothesis has been triggered by the result of several 
recent investigations \cite{Reut98,Dou98,Laus02a,Reut02b,Laus02b,
Soum99,Perc03,Liti04,Bona05},
which support the possibility that Quantum Einstein Gravity (QEG),
the quantum field theory of gravity whose underlying degrees of
freedom are those of the spacetime metric, can be defined
nonperturbatively as a fundamental, ``asymptotically safe''
\cite{Wein79} theory. By definition, its bare action is defined at a 
non--Gaussian renormalization group (RG) fixed point. In the framework
of the effective average action \cite{Berg02,Wett01,Reut94} a suitable
fixed point is known to exist in the Einstein--Hilbert truncation of
theory space \cite{Reut98,Laus02a,Soum99} and in the higher--derivative
generalization \cite{Laus02b}. Detailed analyses of the
reliability of this approximation \cite{Laus02a,Reut02b,Laus02b} and the  
conceptually independent investigations  
\cite{Bona05,Nied03,Nied02,Forg02} suggest that the
fixed point should indeed exist in the exact theory, implying its
nonperturbative renormalizability.

Within this framework, gravitational phenomena at a 
typical distance scale $\ell\equiv k^{-1}$  are described
in terms of a scale-dependent effective action 
$\Gamma_k[g_{\mu\nu}]$ which should be thought of as a Wilsonian
coarse-grained free-energy functional. 
The mass parameter $k$ is an infrared cutoff in the sense that
$\Gamma_k$ encodes the effect of all metric fluctuations with 
momenta larger than $k$,
while those with smaller momenta are not yet ``integrated out''. 
When evaluated at tree level,
$\Gamma_k$ describes all processes involving a single 
characteristic momentum $k$ with all
loop effects included.

In Ref. \cite{Reut98}, $\Gamma_k$ has been identified with the 
effective average action for
Euclidean quantum gravity and an exact functional RG equation 
for the $k$-dependence
of $\Gamma_k$ has been derived. Nonperturbative solutions were 
obtained within the
``Einstein-Hilbert truncation'' which assumes $\Gamma_k$ to be of the form
\be\label{average}
\Gamma_k=
(16\pi G(k))^{-1}\int d^4 x \sqrt{g} \{ -R(g) 
+ 2\Lambda(k)\}.
\ee
The RG equations yield an explicit answer for the 
$k$-dependence of the running Newton term
$G(k)$ and the running cosmological term $\Lambda(k)$. 
In Refs. \cite{Bona02b} and \cite{Bona99,Bona00} it was argued that 
they are important for an understanding of the Planck 
era immediately after the big bang and the structure of black hole singularity.
However, there are indications \cite{Tsam93} that quantum 
Einstein gravity, because of its
inherent IR divergences, is subject to strong renormalization 
effects also at very large distances.
In cosmology those effects would be relevant for the Universe at 
late times, and it has been speculated
that they might lead to a dynamical relaxation of $\Lambda$, thus 
solving the cosmological
constant problem \cite{Tsam93}.  
  
In Ref. \cite{Bona02a}  it was then argued that the 
late expansion of the Universe can be viewed as a renormalization group 
evolution near a non-Gaussian infrared fixed point 
(hereafter IRFP) where 
$G$ and $\Lambda$ become running quantities at some late time. 
In that  work a sharp transition between standard FLRW cosmology and
accelerated RG driven expansion was supposed to occur at some time, but  
a precise link with an underlying renormalization 
group trajectory was still missing, so  that the FP was reached 
exactly at the transition. 
In spite of this simplifying assumption,
the agreement with several high redshift 
observations is impressive \cite{Bent04}.
The aim of this paper is to improve the model described 
in Refs. \cite{Bona02a,Bent04} by taking into account the scaling evolution
near the IRFP.  In particular, we shall show that it is  possible 
to describe the accelerated expansion of the Universe 
as a scaling evolution into an {\it attraction basin} 
of the IRFP. Our approach is in principle very similar to the standard 
procedure used in statistical mechanics  based on the linearization
of the RG equation near a fixed point, although we do not need    
to explicitly solve the RG equations. 
As in the case of QCD before the discovery of asymptotic freedom,  
it is  possible to 
constrain the behaviour of the Gell--Mann--Low function $\beta (g)$  
by means of the properties of the linearized RG flow near a 
fixed point, without explicitly solving the RG
equations \cite{Wils71}. 
For gravity it is  possible  to explicitly calculate 
the critical exponents 
by imposing that an allowed RG flow should be dynamically consistent with the 
Bianchi integrability condition. It turns out that the only possible solution 
predicts that the Universe approaches the IRFP 
along one irrelevant and one marginal direction.

Section 2 studies the RG-improved Einstein equations when the 
energy-momentum tensor takes the perfect-fluid form. Section 3 studies
a Bianchi I model with a multicomponent fluid, while concluding remarks 
and a critical assessment are presented in section 4. 

\section{RG-improved Einstein equations}

In the following we shall present the improved 
RG equation in the $3+1$ formalism. 
Let $g_{\mu\nu}$ be the space-time metric with signature $(-,+,+,+)$. 
A ``cosmological fundamental observer'' comoving with the cosmological fluid
has 4-velocity $u^\mu= dx^\mu/ d\tau$ with $ u^\mu u_\mu = -1$,
where $\tau$ is the proper time along the fluid flow lines.
The projection tensor onto the tangent 3-space orthogonal to $u^\mu$ is
$h_{\mu\nu} = g_{\mu\nu}+u_\mu u_\nu$,
with  ${h^\mu}_\nu {h^{\nu}}_\sigma =
{h^\mu}_\sigma$ and ${h^{\mu}}_\nu u^\nu =0$.
We denote by a semicolon the standard covariant derivative 
and by an over-dot the 
differentiation with respect to the proper time $\tau$.
The covariant derivative of $u^{\mu}$ reads as 
\be\label{2.3}
u_{\mu;\nu}=\omega_{\mu\nu}+\sigma_{\mu\nu}+
{1\over 3}\Theta h_{\mu\nu}-\dot{u}_{\mu} u_{\nu},
\ee
where  $\omega_{\mu\nu}={h^{\alpha}}_\mu {h^{\beta}}_\nu u_{[\alpha;\beta]}$
is the vorticity tensor, $\sigma_{\mu\nu}
= {h^{\alpha}}_\mu {h^{\beta}}_\nu u_{(\alpha;\beta)}
-{1\over 3}\Theta h_{\mu\nu}$
is the shear tensor, $\Theta = {u^\mu}_{;\mu}$ is the expansion scalar and
$\dot{u}^\mu = {u^{\mu}}_{;\nu} u^\nu$ is the acceleration 
four-vector; square and round
brackets denote anti-symmetrization and symmetrization, respectively.
The Einstein equations read 
\be\label{2.4}
R_{\mu\nu}-{1\over 2} R g_{\mu\nu}= -\Lambda g_{\mu\nu}+8\pi G
T_{\mu\nu},
\ee
where $\Lambda=\Lambda(x^\mu)$ is the position-dependent 
cosmological term and
$G=G(x^\mu)$ the position-dependent Newton parameter. Note that, unlike
the analysis by Reuter and Weyer \cite{Reut04b}, 
who ``improve'' the action functional,
our right-hand side in Eq. (2.2)  
does not contain covariant derivatives of the
Newton parameter nor a term describing the 4-momentum carried by $G$
and $\Lambda$.

The energy-momentum tensor $T_{\mu\nu}$ is assumed to be covariantly 
conserved. For a perfect fluid it has the form
$T^{\mu\nu} = (p+\rho) \; u^\mu u^\nu + p\;g^{\mu\nu}$.
The conservation law ${T^{\mu\nu}}_{;\nu}=0$ 
leads to mass-energy conservation
\be\label{2.6}
{\dot{\rho}} +\Theta(\rho +p)=0,
\ee
and to the equation of motion
\be\label{2.7}
\dot{u}^\mu + {h^{\mu\nu}p_{;\nu}\over \rho + p} = 0.
\ee
The Bianchi identities require the RHS of Eq. (\ref{2.4}) to be covariantly
conserved. This consistency condition, together with
the conservation laws (\ref{2.6}) and (\ref{2.7}),
provides the equations for $\Lambda$ and $G$, i.e. (see comments below
Eq. (2.13))
\ba\label{2.8}
&&\dot{\Lambda} + 8 \pi\dot{G}\rho = 0, \\[2mm]
&&h^{\mu\nu}\Lambda_{;\nu} - 8\pi p \;h^{\mu\nu} {G}_{;\nu} =0,
\label{2.8b}
\ea
by projecting along $u^\mu$ and onto the hyperplane orthogonal to $u^\mu$.
These equations differ from the consistency condition (2.20) obtained in
Ref. \cite{Reut04b} from an improved action functional with variable
$G$ and $\Lambda$.

The {\it Raychaudhuri equation} is obtained with the help of the Einstein
field equations and of Eq. (\ref{2.3}), i.e.
\be\label{2.9}
\dot{\Theta}+{1\over 3}\Theta^2+2(\sigma^2-\omega^2)-
\dot{u}^\mu_{\;\; ;\mu}+4 \pi G (\rho+3p)-\Lambda=0,
\ee
where $2\sigma^2 \equiv  \sigma_{\mu\nu}\sigma^{\mu\nu}$ and
$2\omega^2 \equiv \omega_{\mu\nu}\omega^{\mu\nu}$. 
The term $ \dot{u}^\mu_{\;\; ;\mu}$ 
is identically vanishing for homogeneous spaces.
The scalar curvature of the tangent space is given by
\be\label{2.10.0}
{\cal K}\equiv {}^{(3)}R = R+2R_{\mu\nu}u^{\mu}u^{\nu}
+2\sigma^2-2\omega^2-{2\over 3}\Theta^{2},
\ee
which leads, by using the field equations (\ref{2.4}), 
to the {\it Friedmann equation}
\be\label{2.10}
{\cal K}= 2\sigma^2 -2\omega^2 -{2\over 3}\Theta^2
+16\pi G\rho +2\Lambda.
\ee
In homogeneous spaces, Eqs. (\ref{2.7}) and (\ref{2.8b}) 
are identically satisfied, while 
the Friedmann equation (\ref{2.10}), the energy  
conservation equation (\ref{2.6}) 
and the integrability condition (\ref{2.8}) constitute the  
evolution equations for kinematical quantities.
In order to integrate them, the evolution equations 
for shear and vorticity are needed, together with 
the dynamical equations for $G$ and $\Lambda$ which are obtained  
by the RG equations. 
The latter are obtained in the Einstein--Hilbert truncation as 
a set of $\beta$-functions
for the dimensionless Newton constant and cosmological constant, 
$g$ and $\lambda$,
\be\label{rg}
k{ \partial_k g } = \beta_g(g,\lambda), \;\;\;\;\;\;\;\;\;\;\;\;\; 
k{ \partial_k \lambda } = \beta_{\lambda}(g,\lambda), \;\;\;\;\;\;
\ee
and the link with the spacetime dynamics is provided by
the so-called {\it cut-off identification}
\be\label{ide1}
k=k(\tau,\rho,\dot{\rho},\Theta, \dot{\Theta},...).
\ee  
The dots stand for all possible physical or geometrical invariants 
which can act as IR 
regulators in the fluctuation determinant of $\Gamma_k$. 
 
The knowledge of the precise functional dependence in Eq. (\ref{ide1})
would then provide a dynamical evolution which is consistent  
with the full effective action at $k=0$. As it was explained in Ref. 
\cite{Bona02a} the simple choice $k\propto 1/t$ can be justified on the
ground that, if there are no other scales in the system when the Universe had
age $t$, fluctuations with frequency greater than $1/t$ 
may not have played any role as yet, 
and the running must be stopped at $k\propto 1/t$.
On the other hand, in the radiative era in the early Universe, 
the energy density can be a better candidate to cutoff the modes
\cite{Gube03}, as is also suggested from general arguments based on
the holographic principle \cite{Horv04}.

At the fixed point, $k$ in Eq. (\ref{ide1}) is entirely 
determined by dimensional analysis, 
and we must always have a power law scaling of the type  
$k \propto 1/\tau$,  $k\propto \rho^{1/4}$ or 
$k  \propto \Theta$, for instance.  
However, for a non-Gaussian fixed point
\be\label{ngfp}
G = \frac{\gira} {k^{2}},\;\;\;\;\;\;\; 
\Lambda = \lira k^{2},
\ee
so that $G\Lambda =\ngf$ and from Eqs. (\ref{2.8}) and (\ref{2.8b})  
a scaling of the type 
\be\label{ide}
k = \xi^{-1/2} \rho^{1/4},
\ee
with $\xi=\sqrt{\lira/8\pi\gira}$ must always hold if  
no spatial variation of $G$ and $\Lambda$ occurs, {\it i.e.}
$h^{\mu\nu}\Lambda_{;\nu}=h^{\mu\nu} {G}_{;\nu} \equiv 0$.

It should be stressed that our analysis relies heavily on the assumption
of a classical time evolution (2.5) in the neighbourhood of the non-trivial
infrared fixed point. Such an assumption limits the number of relevant
operators that would otherwise contribute to (2.5). Hence the scaling in
Eq. (2.13) derives from the {\it ad hoc} assumption (2.5). It remains
to be seen whether such an approximation is appropriate for studying
the very late universe (see further comments at the end of section 4).

In particular, Eq. (\ref{ide}) will be satisfied in any homogeneous cosmology,
as already noticed by Ref. \cite{Babi02} for a class of FLRW universe models, 
and for any matter field,  
so that it is always possible to close the system of the 
RG improved Einstein equations in a mathematically 
consistent way in these cases\footnote{ 
Note that there is no need to further invert the relation 
(\ref{ide}) with $\rho=\rho(t)$ to determine $k=k(t)$, since 
this map is not in general one-to-one   
if other dynamical and/or geometrical scales occur in the system 
\cite{Babi02}.}. 
On the other hand, we shall see that the identification 
$k\propto 1/t$ is always recovered from Eq.
(\ref{ide}) in the limit of $t$ much larger than all relevant mass 
scales in the system.

\section{Bianchi I model}

Let us now consider a Bianchi I model with a multicomponent fluid.
The line element reads as 
\be
ds^2 = -dt^2 +a_{1}^{2}(t)dx^2+a_{2}^{2}(t)dy^2 +a_{3}^{2}(t)dz^{2},
\ee
and in this case we have three different expansion factors. During the
initial early universe stage, the dominant contribution in the Raychaudhuri 
Eq. (\ref{2.9}) is the shear, and it is not {\it a priori} clear 
which the most physically plausible cutoff could be.
On the other hand, the evolution equations described before 
are completely covariant, and Eq. 
(\ref{ide}) still holds. The scalar curvature ${\cal K}$ is identically zero 
in Eq. (2.8); moreover, by virtue of Eqs. (2.1)--(2.3), 
(2.5), (2.7) and (2.8) 
the shear scalar is found to obey, even with variable $G$
and $\Lambda$, the relation 
\be
\sigma^2 = \Sigma/S^{6},
\ee
$\Sigma$ being a non-negative real number, and the     
scale $S(t)$ being defined as $S \equiv (a_1a_2a_3)^{1/3}$.   
The expansion scalar is then given by
$\Theta=3\dot{S(t)}/ S(t)\equiv 3H$, $H$ being the global
Hubble parameter. Let us further consider a 
two-component perfect fluid described by
$\rho = \rho_1+\rho_2$ and $p_1 = w_1 \rho_1$, $p_2= w_2 \rho_2$, 
so that, from the conservation law (\ref{2.6}),
we immediately get
\be\label{cons}
\rho = \rho_1+\rho_2 = {\mt_1 \over 8\pi S^{3+3w_1}}+
 {\mt_2 \over 8\pi S^{3+3w_2}},
\ee
$\mt_1$ and $\mt_2$ being integration constants 
(the factor $8\pi$ has been inserted for convenience).
On inserting Eqs. (\ref{ngfp}), (\ref{ide}) and (\ref{cons}) 
into Eq. (\ref{2.10})
we obtain a single differential equation for $S(t)$, i.e.  
\be\label{newf}
\dot{S}^{2} 
={2\over 3}\sqrt{\ngf}\Big ({\mt_1 \over  S^{3w_1-1}}+
{\mt_2 \over  S^{3w_2-1}}\Big)^{1/2}+{\Sigma\over 3 S^4}.
\ee

It is interesting to discuss two particular cases, i.e.
a mixture of matter and radiation with no shear 
(the usual FLRW universe), and 
a stiff matter dominated universe, with shear. In the former case
the new scale is represented by the energy density of the second fluid 
component. In the latter case
the new scale present in the system is the shear. 

If we thus set $w_1=1/3, w_2=0, \Sigma=0$ 
in Eq. (\ref{newf}), and the solution with 
$S(0)=0$ is given by
\be\label{sol1}
S(t)= {1\over \mt_2}\Big [ \mt_2\sqrt{{3 \over 8}\sqrt{\gira\lira} }
(t+t_c )\Big ]^{4/3}-{\mt_1\over \mt_2},
\ee
with $t_c= \sqrt{{8/ 3 \sqrt{\lira\gira}}} 
\mt_1^{3/4}/\mt_2$.
If instead $\mt_2=0$ and $w_1=1$ (stiff matter), we obtain
\be\label{sol2}
S(t)= \Big [ {3\over 2 \sqrt{\lira\gira\mt} }
\Big ( \sqrt{\gira\lira\mt} t 
+{\Sigma\over \sqrt{3}}\Big )^2 
-{\Sigma^2 \over 2  \sqrt{\lira\gira\mt}} \Big ]^{1/3}
\ee
for a non-vanishing shear.
Note that, in this latter case, the system becomes isotropic 
at late times because
\be
{\sigma \over\Theta}  = {\Sigma \over 3 
\sqrt{\lira\gira \mt}t+\Sigma \sqrt{3}}.
\ee
In both cases the cutoff identification in terms 
of the cosmic time $t$ is recovered in the 
limit $t\gg t_c$ or $t\gg \Sigma /\sqrt{\mt}$
(as can be seen by direct substitution of 
Eq. (\ref{sol1}) or Eq. (\ref{sol2}) in Eq. (\ref{cons})
and by further taking this limit),
which are the additional dimensionful scales of the system, 
but the cutoff identification in Eq. (\ref{ide})  
with $\xi= \sqrt{\lira /8\pi \gira}$ is realized at any time. 
This result suggests that this type of identification (with possibly
a different value of $\xi$) is more powerful than the choice $t \propto 1/k$, 
and it can possibly be used also 
in the attraction basin of the fixed point, as we shall discuss shortly. 

In fact the IRFP hypothesis described in Refs. \cite{Bona02a,Bent04} 
assumes that the fixed point is non-Gaussian and attractive.
We can thus describe its attraction basin 
by means of a two-dimensional subspace of irrelevant (or marginal) operators.
Quite generally we can write 
$g(k)= \gira +\gu k^{\theta_1}$ 
and $\lambda(k) = \lira + \lu k^{\theta_2}$, where 
$\theta_1 \geq0 $ and $\theta_2\geq 0$ are the critical exponents.
The dimensionful $G$ and $\Lambda$, by virtue of Eq. (\ref{ide}),  
read as 
\be\label{glb}
G={\xi\over \sqrt{\rho}}\Big( \gira 
+ \gu\xi^{-\theta_1/2}\rho^{\theta_1/4}\Big), 
\;\;\;\;\;\;\;
\Lambda = {\sqrt{\rho}\over \xi}\Big (\lira 
+\lu\xi^{-\theta_2/2}\rho^{\theta_2/4}\Big).
\ee
By inserting Eqs. (\ref{glb}) in the integrability 
condition (\ref{2.8}),  we readily find that the only solution
valid for any value of the scaling eigenoperators $h_1$, $h_2$
is obtained with $\xi= \sqrt{{(\lira +\lu) / 8\pi \gira}}$,
and the  {\it critical exponents of the Universe} 
are given by $\theta_1 =2$ and $\theta_2 =0$.
In other words, the IRFP is approached along one irrelevant 
and one marginal direction.  

From the Friedmann 
equation (\ref{2.10}), in the case of $\Sigma=0$ and $\mt_2=0$, 
we find that the time evolution of the scale factor is then ruled by 
\be\label{newf2}
\dot{S}^2(t)= {2\over 3}{\sqrt{\cchi\mt}\over S^{(3w-1)/2}}
+{\gu\mt \over  S^{3w+1}}.
\ee
The solution for a dust dominated Universe $w=0$ is thus given by 
\be\label{solf}
S(t) =  \Big [{3 \sqrt{\cchi\mt} \over 8 }\Big 
(t+\sqrt{ {4\gu\over 3 {\cchi}}} \Big)^2 
-{\gu \over 2}\sqrt{{\mt \over \cchi}} \Big ]^{2/3},
\ee 
which approaches the $S\propto t^{4/3}$ behaviour found 
in Ref. \cite{Bona02a} for large times, 
but reproduces the standard $S\propto t^{2/3}$ for 
times much smaller than the 
{\it crossover time} $t_c \sim \sqrt{\gu}$. 
For $t>t_c$ the Universe actually enters 
the attraction basin of the IRFP, which is eventually
reached for $t=\infty$.
The deceleration parameter is given by 
\be\label{qq}
q= - \frac{3(3\cchi t^2+4 \sqrt{3 \cchi h_1}t-8h_1)}{4(3\sqrt{\cchi}t
+2\sqrt{3 h_1})^2},
\ee
which tends to $-1/4 $ for $ t\gg t_c$, where  
$\Omega_m=\Omega_{\Lambda}=1/2$  as in the original 
IRFP model \cite{Bona02a}.

The last question we would like to answer is 
how to justify the domain of validity of our 
scale identification $k\propto \rho^{1/4}$ in terms of a 
cutoff function. More precisely, 
we would like to understand whether the use of the energy density 
as a cutoff can be understood in
terms of suppression of the fluctuations at a scale 
$k\propto \rho^{1/4}$. Let us consider, for the sake of simplicity,
a generic self-interacting scalar theory.  
If we denote by $V(\phi)$ the scalar field potential, the energy density 
reads as $\rho = \dot{\phi}^2/2+V(\phi)$. 
On the other hand, the contribution to the fluctuation
determinant in the matter sector can easily be estimated in the 
proper-time formulation of Ref. \cite{Bona05},
where the cutoff is realized essentially by the spectrum of the second 
functional derivative of the action. 
In this case the modes are cut-off at a nonvanishing 
mass scale $k \propto \sqrt{V''(\phi)}$, the prime denoting the 
functional derivative with respect to $\phi$. 
In order for the energy density to represent a meaningful cutoff, it must 
satisfy $\rho \propto k^4 \propto V''^2$. This relation can be satisfied 
if the scalar-field evolution is dominated by the potential term. 
In fact in this case (slow-roll approximation), the 
kinetic term is negligible as compared to the 
potential term and one must have $\rho \propto V \propto V''^2$, 
which is always realized for a familiar self-interacting 
scalar theory of the type $V \propto \phi^4$. This result suggests that, 
if the dynamical evolution is dominated by the potential term,      
a scaling of the  type $k \propto \rho^{1/4}$ encodes the 
relevant degrees of freedom
whose fluctuations of momenta greater than $k$ are suppressed. 
Of course, this approximation 
becomes increasingly reliable in the infrared region, where only low-momentum
modes are taken into account.

\section{Concluding remarks}

We have derived a smooth transition between 
standard cosmology and the IRFP modified cosmology 
described in Ref. \cite{Bona02a} 
which uses the scaling $k \propto \rho^{1/4}$ as a cut-off identification.
We find, for the first time in the literature, that
the dynamical evolution in the neighbourhood of the non-Gaussian 
infrared fixed point can be described
by means of one marginal and one irrelevant direction as the only possible
RG trajectory which is consistent with the Bianchi identities.     
It would be therefore interesting to
test this model against the SnIa data, and we plan to 
investigate this issue in a separate work. 

At present, a substantial amount of literature, with detailed calculations,
supports non-perturbative renormalizability of quantum Einstein gravity,
by virtue of an ultraviolet cutoff at a non-Gaussian renormalization
group fixed point (see Ref. \cite{Laus05} 
and references therein). Hopefully, the
years to come will provide at least a toy-model derivation of the improved
Einstein equations (2.2) with running to an infrared fixed point. Until this
remains an open problem, our original calculations are only a promising 
indication of the potentialities of the approach advocated.

For the purpose of testing the infrared fixed-point hypothesis, the
improved action functional built by some of us in Ref. \cite{Bona04} might
also prove useful. At a deeper level, the main problem one faces is 
as follows \cite{Reut04b}: simple local truncations are sufficient in tre
ultraviolet, but for $k \rightarrow 0$ non-local terms should be included
in the truncation ansatz for $\Gamma_{k}$. Although we have still used a
strictly local truncation, the cutoff identification $k=k(x)$ introduces
non-local features into the theory which, under certain conditions, are
equivalent to some of the non-local terms in the truncated effective
average action $\Gamma_{k}$ (cf. Eq. (1.1)). We refer the reader to the
Introduction of Ref. \cite{Reut04b} for a discussion of the partial
equivalence.

It will be also interesting to compare our accelerated expansion of the 
universe with the one recently obtained within the framework of $f(R)$
theories (see, for example, Ref. \cite{Cogn05}). As far as the general 
theoretical background is concerned, we refer the reader to
Ref. \cite{Buch92} for effective-action methods. On the experimental 
side, the variation of $G$ is subject to experimental limits that may
constrain the theory \cite{Will93}, and we hope to be able to discuss
this point as well in future work.
 
\acknowledgments
This work was partially supported by the Marie Curie ``Transfer of Knowledge''
Project no. 002995 COSMOCT-Cosmology and Computational Astrophysics at
Catania Astrophysical Observatory, under the VI R \& D Framework Programme
of the European Commission.
AB would like to thank M. Reuter for important comments on the manuscript. 
The authors are grateful to the INFN for financial support. The work of
G. Esposito has been partially supported by PRIN {\it SINTESI}.

\end{document}